\documentclass[11pt]{JHEP3}

\usepackage{amsfonts}
\usepackage{amssymb}
\usepackage{amsmath}
\usepackage{mathrsfs}
\usepackage{latexsym}
\usepackage{bm}
\usepackage{bbm}
\usepackage[mathcal]{euscript}
\usepackage[dvips]{epsfig}
\usepackage{epsf}
\usepackage{feynarts}

\title{A theoretical determination of the $\eta-\eta'$ mixing}
\author{C\'eline Degrande and Jean-Marc G\'erard \\
{\small\it Centre for Particle Physics and Phenomenology (CP3),} \\
{\small\it Universit\'e catholique de Louvain,} \\
{\small\it Chemin du Cyclotron, 2, B-1348, Louvain-la-Neuve, Belgium.}\\
E-mail: \email{celine.degrande@uclouvain.be}, \email{jean-marc.gerard@uclouvain.be}}

\abstract{The systematic large $N_c$ limit within chiral perturbation provides an optimal $\eta-\eta'$ mixing angle of about -27$^\circ$ at leading order in $p^2$. In this frame, agreement with the data can be reached with higher order corrections of about 20\%.}


\preprint{}

\begin{document}

\section{Introduction}

\paragraph{}A vast literature on phenomenological descriptions of the $\eta-\eta'$ system has been published in the past ten years \cite{FEF}. Yet, the $\eta-\eta'$ mixing angle alone is more than an effective parameter to be extracted from low energy data. Its peculiar value may indeed shed some light on the non-perturbative dynamics of the fundamental QCD theory and in particular on the axial $U(1)$ anomaly. Needless to recall here why the subsequent parity (P) and time-reversal (T) violations constitute a major puzzle in the Standard Model for electroweak and strong interactions \cite{cernsc}.

\paragraph{}To link this axial anomaly with the observed mass spectrum for the pseudoscalar meson nonet, alternative paths based on the chiral perturbation theory or the large number of colours limit have been proposed. Among them, the chiral perturbation theory at leading order in $p^2$ \emph{and} $1/N_c$ proves efficient once the typical 20\% corrections expected from the flavour symmetry breaking are duly acknowledged. 
\paragraph{}Within this rather simple framework, the $\eta(\eta')$ masses are functions of the mixing angle $\theta$. In particular, the $\eta-\eta'$ mass ratio is not fixed by the theory but can only be optimized with respect to its experimental value for $\theta\approx-27^\circ$. However, the corrections requisite to reproduce the measured value of this ratio raise the question of the systematic expansion to adopt. It appears that including the next to leading order in $p^2$ in the large $N_c$ limit is quite predictive and compatible with the data. Consequently, this approach requires the $1/N_c$-suppressed one-loop contributions to be small. In this Letter, we emphasize that the optimal value of the $\eta-\eta'$ mixing angle at leading order turns out to consistently damp out the quadratically divergent one-loop corrections to the $\eta-\eta'$ inverse propagator matrix and the $\eta'\rightarrow \eta\pi\pi$ decay amplitude.

\section{An effective theory at leading order in $p^2$ and $\frac{1}{N_c}$}

\paragraph{}If $n$ quark flavours are massless, the fundamental Lagrangian of QCD displays a global $U(n)_L\otimes U(n)_R$ invariance. In the large $N_c$ limit, $N_c$ being the number of colours, the effective Lagrangian which features this chiral symmetry at lowest order in $p^2$  reads \cite{witten}
\begin{equation}
 \mathscr{L}^{\left(p^2,0\right)} = \frac{f^2}{8}\left[\left<\partial_\mu U\partial^\mu U^\dagger\right>+r\left<mU^\dagger+Um^\dagger\right>\right]\label{lagrange}
\end{equation}
where $U$ is a $n$-by-$n$ matrix transforming as $U\rightarrow g_L U g_R^\dagger$. The mass matrix $m$ for the light quarks transforms as $U$ and its determinant is assumed to be real to ensure P and T invariance. In Eq.\eqref{lagrange}, the parameters with dimensions of mass scale respectively as
\begin{equation}
 f\varpropto N_c^{1/2},\qquad r\varpropto N_c^0.
\end{equation}
 In the large $N_c$ limit, $U(n)_L\otimes U(n)_R$ has to be spontaneously broken into the maximal vectorial subgroup $U(n)_V$ if $n\geq3$ \cite{CW}. Consequently, $U$ is a unitary field which can be expanded around its vacuum expectation value as a function of the Goldstone bosons nonet, 
\begin{equation}
 U = \mathbbm{1} + i\sqrt{2}\frac{\pi}{f} - \frac{\pi^2}{f^2} + \mathscr{O}(\frac{\pi^3}{f^3}).
\end{equation}
In the case of three light flavours,  
\begin{eqnarray}
 \pi &=& \left(\begin{array}{ccc}\pi^3+\frac{1}{\sqrt{3}}\eta^8+\sqrt{\frac{2}{3}}\eta^0&\sqrt{2}\pi^+&\sqrt{2}K^+\\\sqrt{2}\pi^-&-\pi^3+\frac{1}{\sqrt{3}}\eta^8+\sqrt{\frac{2}{3}}\eta^0&\sqrt{2}K^0\\\sqrt{2}K^-&\sqrt{2}\,\overline{K^0}&-\frac{2}{\sqrt{3}}\eta^8+\sqrt{\frac{2}{3}}\eta^0
\end{array}\right)
\end{eqnarray}
and the masses of the pseudoscalars can be easily extracted once $m$ is diagonalized. Working from now in the isospin limit $m_u=m_d=\tilde{m}$, we obtain
\begin{eqnarray}
 m_\pi^2 &=& r\tilde{m}\\
 m_K^2 &=& \frac{r}{2}\left(\tilde{m}+m_s\right)
\end{eqnarray}
and  
\begin{equation}
m_{8-0}^2 = \frac{1}{3}\left(\begin{array}{cc}
4m_K^2-m_\pi^2&-2\sqrt{2}\left(m_K^2-m_\pi^2\right)\\-2\sqrt{2}\left(m_K^2-m_\pi^2\right)&2m_K^2+m_\pi^2
\end{array}
\right)  \label{massm}
\end{equation}
with the octet-singlet flavour basis conventionally characterized by the following amount of strange/non-strange quarks  in the meson wave function
\begin{eqnarray}
 \eta^8 &\sim& \frac{1}{\sqrt{6}}\left(u\bar{u}+d\bar{d}-2s\bar{s}\right)\\
 \eta^0 &\sim& \frac{1}{\sqrt{3}}\left(u\bar{u}+d\bar{d}+s\bar{s}\right).
\end{eqnarray}
At this level, the masses of the physical pseudoscalar fields
\begin{eqnarray}
 \left(\begin{array}{c}\eta\\\eta'\end{array}\right)&=&\left(\begin{array}{cc}\cos\theta&-\sin\theta\\\sin\theta&\cos\theta 
\end{array}\right)\left(\begin{array}{c}\eta^8\\\eta^0\end{array}\right)
\end{eqnarray}
are only functions of the $\pi$ and $K$ ones and vanish in the chiral limit \mbox{$m_u=m_d=m_s=0$}. However, the measured mass of the $\eta'$ around 1 GeV tells us that the axial $U(1)$ has been broken by the dynamics of QCD itself \cite{thooft}. In the limit of a large number of colours within chiral perturbation, this explicit breaking is implemented through the one and only term \cite{witten} 
\begin{equation}
 \mathscr{L}^{\left(p^0,1/N_c\right)} = \frac{f^2}{8}\frac{m_0^2}{4N_c}\left<\ln U-\ln U^\dagger\right>^2=-\frac{1}{2}m_0^2\eta_0^2+\mathscr{O}\left(\pi^4\right)\label{ano}
\end{equation}
which is $1/N_c$-suppressed but $p^0$-enhanced with regard to the effective Lagrangian \eqref{lagrange}. Accordingly, the $\eta_0-\eta_0$ element $m_{00}^2$ of the mass matrix \eqref{massm} is corrected by the parameter $m_0^2$ so that the $\eta,\,\eta'$ masses are not anymore fixed in terms of the $\pi$ and $K$ masses but are functions of the mixing angle $\theta$, as displayed in Fig.\ref{figMasses}:
\begin{eqnarray}
 m_\eta^2 &=& \frac{1}{3}\left[4m_K^2-m_\pi^2+2\sqrt{2}\left(m_K^2-m_\pi^2\right)\tan\theta\right]\label{massee}\\
 m_{\eta'}^2 &=& \frac{1}{3}\left[4m_K^2-m_\pi^2-2\sqrt{2}\left(m_K^2-m_\pi^2\right)\cot\theta\right]\label{masseep}.
\end{eqnarray}
The resulting relation between physical quantities defined at lowest order
\begin{equation}
 \tan^2\theta = \frac{m_\eta^2-\frac{1}{3}\left(4m_K^2-m_\pi^2\right)}{\frac{1}{3}\left(4m_K^2-m_\pi^2\right)-m_{\eta'}^2}\qquad\left(\left|\theta\right| = 11.4^\circ\right)\label{GMO}
\end{equation}
is analogous to
\begin{equation}
 \tan^2\theta_W = \frac{m_Z^2-m_W^2}{m_W^2-m_\gamma^2}\,\,\,\qquad\qquad\left(\left|\theta_W\right| = 28.2^\circ\right).\label{smangle}
\end{equation}
In other words, the Gell-Mann-Okubo (GMO) mass relation $m_{88}^2=\frac{1}{3}\left(4m_K^2-m_\pi^2\right)$ in the $\eta_8-\eta_0$ mass matrix \eqref{massm} plays here the role of the isospin mass relation $m_{W_3}^2=m_{W^\pm}^2$ in the $W_3-B_0$ mass matrix of the Standard Model for electroweak interactions. The latter relation is known to be invariant under the unbroken custodial SO(3) of the Higgs potential; the former is invariant under the unbroken vectorial $SU(2)_I\otimes U(1)_Y$ since the quark mass matrix $m$ in Eq.\eqref{lagrange} transforms at most as a singlet and an octet of $SU(3)_V$. A breaking of the GMO relation for $m_{88}^2$ would require $\mathscr{O}\left(p^4,0\right)$ terms like $\left<mU^\dagger mU^\dagger\right>$ with $m\otimes m$ also transforming as a \underline{27} under the vectorial flavour group. 
\paragraph{}Surprisingly, even with the additional parameter $m_0^2$, the masses of $\eta$ and $\eta'$ cannot be fitted simultaneously \cite{georgi}. Indeed, taking away $m_K^2$ from Eqs.(\ref{massee}-\ref{masseep}), we easily obtain
\begin{eqnarray}
\frac{m_\eta^2-m_\pi^2}{m_{\eta'}^2-m_\pi^2}&=& \tan\left(2\theta_{th}-\theta\right)\tan\theta\quad\left(\tan2\theta_{th}\equiv-\sqrt2\right)\nonumber\\&\leq&\tan^2\theta_{th}=2-\sqrt{3}.\label{bound}
\end{eqnarray}
In the safe $m_\pi^2\rightarrow0$ limit, the resulting upper bound of $0.27$ for the $\eta-\eta'$ square mass ratio is clearly at variance with the corresponding experimental value of about 0.33. 

\EPSFIGURE[h]{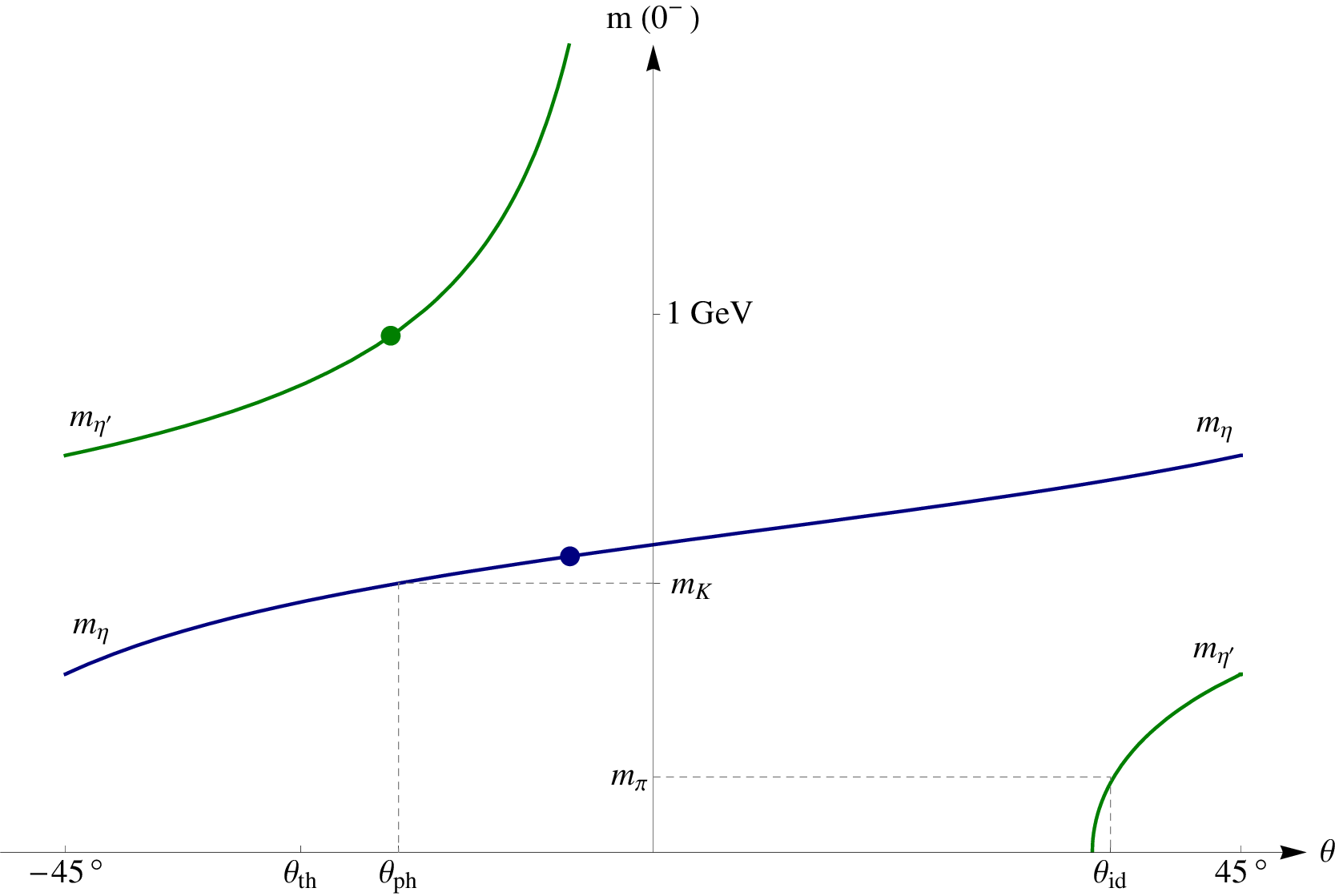, width = 12cm}{The $\eta$ and $\eta'$ masses as a function of their mixing angle from Eqs \eqref{massee} and \eqref{masseep}. We choose to work with $\theta\in\left[-\frac{\pi}{4},\,+\frac{\pi}{4}\right]$  to avoid the renaming $\eta\rightarrow\eta',\,\eta'\rightarrow-\eta$ at $\theta=-\frac{\pi}{4}$. If $m_{\pi,K}$ are fixed at their experimental values, the measured $\eta$ and $\eta'$ masses denoted by dots cannot be simultaneously reproduced at lowest order.\label{figMasses}}

\paragraph{}Mass corrections of about 20\%, as requested by Eq.\eqref{bound} to reproduce the observed $\eta-\eta'$ spectrum, drastically change the absolute value of the mixing angle derived in Eq.\eqref{GMO}. Indeed the physical mass of the $\eta$ and the octet mass $m_{88}$ turn out to be numerically close, within a few percent. Therefore, any departure of lowest order $\eta$ mass from its physical value is enough to produce a major modification of the angle $\theta$ extracted with the help of Eq.\eqref{massee}, as illustrated in Fig.\ref{figMasses}. So, a determination of the mixing angle at lowest order is sensible only if its value is stable with regard to $1/N_c$ and chiral corrections. In this respect, any enlarged symmetry beyond the custodial one is welcome to tame the quantum corrections. For example, a parity-conserving local $SU(3)_L\otimes SU(3)_R$ extension of the $SU(2)_L\otimes U(1)_Y$ electroweak gauge symmetry \cite{unif} covers the custodial $SO(3)$ and would imply
\begin{equation}
\tan\theta_W=-\frac{1}{\sqrt{3}} \qquad\left(\theta_W=-30^\circ\right)\label{gut}
\end{equation}
in pretty good agreement with the on-shell absolute value of the weak mixing angle already introduced in Eq.\eqref{smangle}. 

\paragraph{}In Eq.\eqref{lagrange}, the canonical kinetic term for the $\pi$ field has a global $SO(9)$ invariance. Both the vectorial $SU(3)$-breaking in Eq.\eqref{lagrange} and the axial $U(1)$-breaking in Eq.\eqref{ano} already violate this symmetry at the level of the terms quadratic in the meson fields. Yet, for particular values of the angle $\theta$, remnants of $SO(9)$ may survive at this level; they correspond to the two mass degeneracies displayed with dashes in Fig.\ref{figMasses}:
\begin{itemize}
 \item If $\theta = \theta_{id}$ with
 \begin{equation}
 \tan\theta_{id} \equiv \frac{1}{\sqrt{2}}\quad\left(\theta_{id}=+35.3^\circ\right),\label{angleId}
 \end{equation}
 the physical $\eta'\sim\frac{1}{\sqrt{2}}\left(u\bar{u}+d\bar{d}\right)$ is degenerate in mass with the pions \cite{weinberg} while $\eta\sim-s\bar{s}$. Note that the negative value $\theta_{id}=-54.7^\circ$ corresponding to the other convention with the $s\bar{s}$ component singled out, namely $\eta\sim\frac{1}{\sqrt{2}}\left(u\bar{u}+d\bar{d}\right)$ and $\eta'\sim+s\bar{s}$, is outside the interval $\left[-\frac{\pi}{4},\,+\frac{\pi}{4}\right]$ (see Fig.\ref{figMasses}). The ideal mixing obtained from Eq.\eqref{massm}, i.e., for $m_0^2=0$, is relevant for the vector meson mass spectrum on which the axial $U(1)$ anomaly has no effect, but totally unrealistic for the pseudoscalar one.
 \item If $\theta=\theta_{ph}$ with
 \begin{equation}
 \tan\theta_{ph} \equiv \frac{-1}{2\sqrt{2}}\quad\left(\theta_{ph}=-19.5^\circ\right),\label{anglePh}
 \end{equation}
 the physical $\eta\sim\frac{1}{\sqrt{3}}\left(u\bar{u}+d\bar{d}-s\bar{s}\right)$ is degenerate in mass with the kaons while $\eta'\sim\frac{1}{\sqrt{6}}\left(u\bar{u}+d\bar{d}+2s\bar{s}\right)$. Here, this sensible value for the mixing angle is called phenomenological since it has been extensively used to study hadronic B decays and, in particular, to explain the striking suppression of $B\rightarrow K\eta$ with respect to $B\rightarrow K\eta'$ \cite{lipkin} if penguin diagrams dominate these processes \cite{jmgkou2}. It is also quite popular because the associated quark components are easy to remember and to handle in a phenomenological quark-diagram description of the decay amplitudes according to their $SU(3)$ properties. 
\end{itemize}
\paragraph{}We have no simple mass degeneracy for the case of $\theta_{th}$ already introduced in Eq.\eqref{bound} but note that the three angles of peculiar interest are related through
\begin{equation}
 \tan2\theta_{th} = \tan\left(\theta_{ph}-\theta_{id}\right) \quad\left(\theta_{th}=-27.4^\circ\right)\label{angleRel}
\end{equation}
with, quite incidentally, $\theta_{th}\approx\theta_W$ if the weak mixing angle turns out to be negative as predicted by some unification theory.
\paragraph{}With respect to possible enlarged symmetries covering the custodial $SU(2)_I\otimes U(1)_Y$, we observe that the mass degeneracies $m_{\eta'}=m_\pi$ and $m_\eta=m_K$ correspond to the breaking patterns $SO(9)\rightarrow SO(4)\otimes SO(4)$ and $SO(9)\rightarrow SO(3)\otimes SO(5)$, respectively. These patterns for $\theta_{id}$ and $\theta_{ph}$ can be understood from the fact that $SO(9)$ group  admits $SU(2)\otimes SU(2)\otimes Sp(4)$ or, equivalently, $SO(4)\otimes SO(5)$ as a maximal subgroup \cite{slansky}. However, such enlarged symmetries are explicitly broken at the level of the full effective theory and thus accidental. Consequently, the finite value of the $\theta_{id}$ and $\theta_{ph}$ mixing angles should not be protected against (quadratically) divergent quantum corrections. The fact that the relations \eqref{angleId} and \eqref{anglePh} are not natural can easily be confirmed through the following one-loop computation.

\section{One-loop corrections to the $\eta-\eta'$ inverse propagator matrix}\label{ipm}

\paragraph{}The unification value \eqref{gut} for the observable weak mixing angle $\theta_W$ can most easily be derived by requiring the one-loop fermionic contribution to the $Z-\gamma$ mixing diagram to be finite \cite{branco}. In the same spirit, let us impose the cancellation of the quadratically divergent one-loop corrections to the $\eta-\eta'$ mixing angle $\theta$.
\paragraph{}In order to compute these corrections, we need now to expand $U$ up to the order $\pi^4$,
\begin{equation}
 U=\mathbbm{1}+\sum_{k=1}^\infty a_k\left(i\sqrt{2}\frac{\pi}{f}\right)^k.\label{dev}
\end{equation}
The parameter $a_1$ may be absorbed into the definition of $f$ while the even coefficients are fixed by the unitarity condition \cite{cronin}
\begin{equation}
 a_1=1,\quad a_2=\frac{1}{2},\quad a_3=b,\quad a_4=b-\frac{1}{8},\; \dots
\end{equation}
with $b$ an arbitrary parameter. For $b=\frac{1}{6}$, we recover the standard form 
\begin{equation}
U=\exp\left(i\frac{\sqrt2\pi}{f}\right)                                                                       
\end{equation}
also suited for an octet of pseudoscalars \cite{LK}. But as shown in ref.\cite{cwz}, any other value of b gives rise to the same T matrix when all external lines are put on the mass shell. Yet, one-loop corrections from the kinetic part of the Lagrangian \eqref{lagrange} induce in principle a momentum-dependent $\eta-\eta'$ mixing term which thus has to be taken off-shell. Again by analogy with the scale dependent $Z^0-\gamma$ mixing induced at one-loop in the Standard Model, let us therefore introduce the propagator formalism \cite{renorm}. 

\paragraph{}If we denote by $-iA_{\chi_1\chi_2}(p^2)$ with $\chi_1,\,\chi_2 = \eta,\,\eta'$ the one-loop contributions to the corresponding two point functions, the inverse propagator matrix $\Sigma$ can be parametrized as follows
\begin{eqnarray}
\Sigma_{\eta\eta} &=&\left(1+Z_\eta\right)\left(p^2-m_\eta^2\right)+\delta m_\eta^2-A_{\eta\eta}\left(p^2\right)\nonumber\\
\Sigma_{\eta'\eta'} &=&\left(1+Z_{\eta'}\right)\left(p^2-m_{\eta'}^2\right)+\delta m_{\eta'}^2-A_{\eta'\eta'}\left(p^2\right)\label{propform}\\
\Sigma_{\eta\eta'} &=&\delta m_{\eta\eta'}^2-A_{\eta\eta'}\left(p^2\right).\nonumber
\end{eqnarray}
The last relation in Eq.\eqref{propform} takes into account the fact that $\eta$ and $\eta'$ are decoupled at tree-level, but leaves open the possibility for the one-loop induced mixing to depend on $p^2$. Imposing the normalization of the kinetic part of $\Sigma_{\chi_i\chi_i}$  to be canonical and the physical masses $m_{\chi_i}$ to be the poles of the propagators, we identify
\begin{equation}
 Z_{\chi_i} =  A_{\chi_i\chi_i}'(m_{\chi_i}^2) 
\end{equation}
and 
\begin{equation}
 \delta m_{\chi_i}^2 = A_{\chi_i\chi_i}(m_{\chi_i}^2) 
\end{equation}
where the prime denotes the derivative with respect to $p^2$. From a one-loop computation, we obtain the following quadratic dependences on the ultraviolet momentum cut-off $\Lambda$: 
\begin{eqnarray}
 Z_\eta &=& 3\left[\left(3-20b\right)+\left(4b-1\right)\cos2\theta\right]\frac{\Lambda^2}{\left(4\pi f\right)^2}\nonumber\\
 Z_{\eta'} &=& 3\left[\left(3-20b\right)-\left(4b-1\right)\cos2\theta\right]\frac{\Lambda^2}{\left(4\pi f\right)^2}\label{wavef}
\end{eqnarray}
and
\begin{eqnarray}
 \delta\left(m_\eta^2+m_{\eta'}^2\right) &=& -2\left(2m_K^2+m_\pi^2\right)\frac{\Lambda^2}{\left(4\pi f\right)^2}\nonumber\\
 \delta\left(m_\eta^2m_{\eta'}^2\right) &=& -6m_\pi^2\left(2m_K^2-m_\pi^2\right)\frac{\Lambda^2}{\left(4\pi f\right)^2}\label{dm}
 \end{eqnarray}
 with
 \begin{eqnarray}
 A_{\eta\eta'}\left(p^2\right) &=& \bigg\{\left[3 (4 b-1) p^2 +2(1-8b) m_K^2+2(2 b-1) m_\pi^2\right]\sin2 \theta\nonumber\\&&+4 \sqrt{2} (2 b-1) \left(m_K^2-m_\pi^2\right) \cos2 \theta\bigg\}\frac{\Lambda^2}{\left(4\pi f\right)^2}.\label{mixprop}
\end{eqnarray}
Here, the pseudoscalar masses $m_{K,\pi}$ and the mixing angle $\theta$ are parameters associated with the lowest order Lagrangian defined by Eqs \eqref{lagrange} and \eqref{ano}. In particular, $m_0^2$ has been taken away with the help of the relation
\begin{equation}
 m_0^2 = \frac{2}{3}\left(1-2\sqrt2 \cot2\theta\right)\left(m_K^2-m_\pi^2\right).\label{tan}
\end{equation}

\paragraph{}In general, the one-loop quadratic divergences can be absorbed by a redefinition of the parameters in the $\mathscr{O}\left(p^2\right)$ Lagrangian. Indeed, the corrections quadratic in the cut-off can be identified with the $d=2$ pole in dimensional regularization. Here, a full cancellation of the $\mathscr{O}\left(p^2,1/N_c\right)$ divergent correction \eqref{mixprop} to the mixing requires 
\begin{equation}
 \tan2\theta\left(p^2\right) = \frac{4\sqrt2\left(2b-1\right)\left(m_K^2-m_\pi^2\right)}{3\left(1-4b\right)p^2+2\left(8b-1\right)m_K^2+2\left(1-2b\right)m_\pi^2}.\label{prop}
\end{equation}
Depending on the parameter b, the mixing angle defined in Eq.\eqref{prop} is not a physical quantity. The only way to get rid of the b-dependence is to choose $p^2=2m_K^2$. At such a momentum consistently located between the $\eta$ and $\eta'$ masses, Eq.\eqref{prop} then provides us with an effective mixing angle $\hat{\theta}$ defined at the QCD scale $m_0^2$ : 
\begin{equation}
 \tan2\hat{\theta}\left(m_0^2\right) = \frac{-2\sqrt2\left(m_K^2-m_\pi^2\right)}{\left(2m_K^2+m_\pi^2\right)}\qquad\left(\hat{\theta}=-25.8^\circ\right).\label{olmix}
\end{equation}
We note that the same expression for an on-shell mixing angle $\theta$ can be obtained by simply fixing $b=\frac{1}{4}$ to cancel the momentum dependence in Eq.\eqref{mixprop}. This value of the parameter b, which suggests the other significant form
\begin{equation}
 U = \frac{\mathbbm{1}+\frac{i\pi}{\sqrt{2}f}}{\mathbbm{1}-\frac{i\pi}{\sqrt{2}f}}
\end{equation}
only suited for a full nonet of pseudoscalars \cite{cronin}, ensures $\theta$-independent wave-function renormalizations, i.e., $Z_\eta=Z_{\eta'}$ in Eq.\eqref{wavef}. As a consequence, the only chiral invariant mass operator that would absorb any divergent $\eta_8-\eta_0$ rotation at $\mathscr{O}\left(p^2,1/N_c\right)$ is proportional to  
\begin{eqnarray}
 \frac{f^2}{16}r\left<mU^\dagger-Um^\dagger\right>\left<\ln U-\ln U^\dagger\right> &=& \left(2m_K^2+m_\pi^2\right)\eta_0^2-2\sqrt2\left(m_K^2-m_\pi^2\right)\eta_0\eta_8\nonumber\\&&+\mathscr{O}\left(\pi^4\right)\label{renormM}
\end{eqnarray}
in full agreement with Eq.\eqref{dm} and Eq.\eqref{olmix}. So, the parity-conserving global $SU(3)_L\otimes SU(3)_R$ plays here the role of the enlarged symmetry which covers the custodial $SU(2)_I\otimes U(1)_Y$. Indeed, Eq.\eqref{renormM} tells us that the chiral symmetry of the full effective theory selects in a natural way one negative value $(\hat{\theta})$ for the $\eta-\eta'$ mixing angle, without spoiling the GMO mass relation for $m_{88}^2$.

\paragraph{}As already anticipated from the explicit breaking of the accidental symmetries $SO(4)\otimes SO(4)$ or $SO(3)\otimes SO(5)$ at the level of terms quartic in the meson fields, neither $\theta_{id}$ nor $\theta_{ph}$ are protected against $\Lambda^2$ quantum corrections. On the contrary, Eq.\eqref{olmix} tells us that the angle $\theta_{th}$ which optimizes the $\eta-\eta'$ mass ratio at lowest order might be natural in the safe limit $m_\pi^2\rightarrow0$. In the fundamental theory (i.e., QCD), the corresponding limit $m_{u,d} \rightarrow 0$ would, in principle, solve the so-called strong CP problem. This rather intriguing link evidently calls for further investigations. 

\section{One-loop corrections to the $\eta'\rightarrow\eta\pi\pi$ decay amplitude}

\paragraph{}For the purpose of computing a $b$-independent one-loop correction involving the $\eta-\eta'$ mixing, let us now consider a physical process with on-shell $\eta$ and  $\eta'$ states.

\subsection{Tree-level amplitude}

\paragraph{}The tree-level amplitude for the $\eta'\rightarrow\eta\pi\pi$ decay reads
\begin{eqnarray}
 A\left(\eta'\rightarrow\eta\pi\pi\right) &=&  \frac{1}{f^2}\left[2\left(2\sqrt2\cos2\theta-\sin2\theta\right)\left(\frac{1}{6}-b\right)\left(m_\eta^2+m_{\eta'}^2+2m_\pi^2\right)\right.\nonumber\\&&\left.+8\left(2\sqrt2\cos2\theta-\sin2\theta\right)\left(b-\frac{1}{8}\right)r\tilde{m}\right.\nonumber\\&&\left.+4\sqrt2\left(\cos2\theta-\sqrt2\sin2\theta\right)\left(b-\frac{1}{6}\right)m_0^2\right]\label{Ampltl}
\end{eqnarray}
where $m_\eta$, $m_{\eta'}$ and $m_\pi$ stand now for the physical masses since they come from the momentum dependence induced by the kinetic term in \eqref{lagrange}. In Eq.\eqref{Ampltl}, the second term proportional to $r$ is due to the mass term in Eq.\eqref{lagrange} and the third one arises from the anomalous part given in Eq.\eqref{ano}. With the help of Eq.\eqref{tan}, we eventually recover the well-known result that the tree-level amplitude 
\begin{equation}
 A\left(\eta'\rightarrow\eta\pi\pi\right) = \frac{m_\pi^2}{3f^2}\left(2\sqrt2\cos2\theta-\sin2\theta\right)
\end{equation}
vanishes if $\theta=\theta_{id}$ and is by far too small to reproduce the measured decay width.

\subsection{One-loop amplitude}

\begin{figure}[h!]
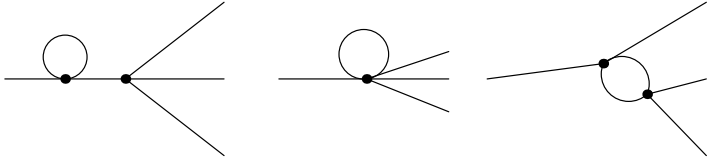

 
\unitlength=1bp%

\begin{feynartspicture}(432,100)(4,1.1)

\FADiagram{}
\FAProp(0.,10.)(5.5,10.)(0.,){/Straight}{0}
\FAProp(20.,17.)(11.,10.)(0.,){/Straight}{0}
\FAProp(20.,10.)(11.,10.)(0.,){/Straight}{0}
\FAProp(20.,3.)(11.,10.)(0.,){/Straight}{0}
\FAProp(5.5,10.)(11.,10.)(0.,){/Straight}{0}
\FAProp(5.5,10.)(5.5,10.)(5.5,14.){/Straight}{0}
\FAVert(5.5,10.){0}
\FAVert(11.,10.){0}

\FADiagram{}
\FAProp(3.,10.)(11.,10.)(0.,){/Straight}{0}
\FAProp(18.5,12.5)(11.,10.)(0.,){/Straight}{0}
\FAProp(18.5,10.)(11.,10.)(0.,){/Straight}{0}
\FAProp(18.5,7.)(11.,10.)(0.,){/Straight}{0}
\FAProp(11.,10.)(11.,10.)(10.5,14.5){/Straight}{0}
\FAVert(11.,10.){0}

\FADiagram{}
\FAProp(0.,10.)(10.6,11.4)(0.,){/Straight}{0}
\FAProp(20.,17.)(10.6,11.4)(0.,){/Straight}{0}
\FAProp(20.,10.)(14.6,8.6)(0.,){/Straight}{0}
\FAProp(20.,3.)(14.6,8.6)(0.,){/Straight}{0}
\FAProp(10.6,11.4)(14.6,8.6)(0.8,){/Straight}{0}
\FAProp(10.6,11.4)(14.6,8.6)(-0.8,){/Straight}{0}
\FAVert(10.6,11.4){0}
\FAVert(14.6,8.6){0}

\end{feynartspicture}
\caption{One-loop topologies for the $\eta'\rightarrow\eta\pi\pi$ decay amplitude.}\label{diag}
\end{figure}

\paragraph{}The one-loop corrections to the process $\eta'\rightarrow\eta\pi\pi$ are associated with the diagrams given in Fig.\ref{diag}. 

The first topology corresponds to the corrections of the inverse propagator given in section \ref{ipm}. The second one involves $\pi^6$ vertices and thus requires the introduction of  the next two coefficients in the development \eqref{dev}, namely
\begin{eqnarray}  
  a_5&=&c\nonumber\\
  a_6&=&c+\frac{b^2}{2}-\frac{b}{2}+\frac{1}{16}.
 \end{eqnarray}
As a result, the $\Lambda^2$-correction to the decay amplitude is given by
\begin{eqnarray}
 \delta A\left(\eta'\rightarrow\eta\pi\pi\right) &=& 4\frac{m_\pi^2}{f^2}\cos^32\theta\bigg[\left(\tan2\theta+\sqrt2\right)\left(\tan^22\theta+\frac{1}{4\sqrt2}\tan2\theta+\frac{1}{2}\right)\nonumber\\&&+\frac{3}{4}\frac{m_\pi^2}{m_K^2-m_\pi^2}\left(\tan2\theta+\frac{1}{2\sqrt2}\right)\tan^22\theta\bigg]\frac{\Lambda^2}{\left(4\pi f\right)^2}.\label{epepp}
\end{eqnarray}
This correction is independent of $b$ and $c$, as it should for any physical quantity, and can be reproduced using the output of FeynRules \cite{feynrules} and Feynarts \cite{feynarts}.

\paragraph{}If we consider again the limit $m_\pi^2\ll m_K^2$, we conclude from Eq.\eqref{epepp} that the optimal value $\theta_{th}$ given in Eq.\eqref{angleRel} for the $\eta-\eta'$ mixing angle indeed damps out the quadratic dependence on the ultra-violet momentum cut-off $\Lambda$, as anticipated from Eq.\eqref{olmix}.

\section{Comments and conclusion}

\paragraph{}In the past, alternative ways to merge the large number of colours limit into the chiral perturbation theory have been used to study the $\eta-\eta'$ system. In particular, the combined expansion
\begin{equation}
 p^2=\mathscr{O}\left(\delta\right), \qquad \frac{1}{N_c}=\mathscr{O}\left(\delta\right)\label{hierarchy}
\end{equation}
has been advocated in ref.\cite{leutwyler}. In this Letter, inspired by the pseudoscalar mass spectrum, we rather follow the approach of ref.\cite{jmgkou} where the leading term in the $1/N_c$ expansion is retained at each order in $p^2$. At the effective level, this implies the hierarchy 
\begin{equation}
 \mathscr{O}\left(p^0,1/N_c\right) > \mathscr{O}\left(p^2,0\right) >\mathscr{O}\left(p^4,0\right),\label{hierar}
\end{equation}
namely
\begin{equation}
 \mathscr{O}\left(p^2,1/N_c\right) \ll \mathscr{O}\left(p^4,0\right)\label{hierar2}
\end{equation}
with the large $N_c$ limit denoted by a zero as in Eqs.\eqref{lagrange} and \eqref{ano}. It amounts to remove the double trace term \eqref{renormM} as well as $\left<\partial_\mu UU^\dagger\right>\left<\partial^\mu U^\dagger U\right>$ in the Lagrangian, \emph{and} to neglect the quadratic one-loop divergences which would renormalize them. The $\eta'\rightarrow\eta\pi\pi$ decay amplitude and the $\eta-\eta'$ mass ratio are known to require sizeable corrections beyond the  $\mathscr{O}\left(p^2,0\right)$ approximation and can thus distinguish between the two working hypothesis \eqref{hierarchy} and \eqref{hierar2}. In ref.\cite{venez} and ref.\cite{georgi}, the $\mathscr{O}\left(p^2,1/N_c\right)$ contributions were invoked for the decay amplitude and the mass ratio, respectively. On the contrary, in ref.\cite{fajfer} and ref.\cite{jmgkou} the $\mathscr{O}\left(p^4,0\right)$ contributions were favoured for these physical quantities, respectively. 

\paragraph{}At $\mathscr{O}\left(p^4,0\right)$, the full set of corrections allows us to naturally reproduce the observed $\eta-\eta'$ mass spectrum. They do not fix by themselves the value of the mixing angle $\theta$ but imply a splitting among the pseudoscalar decay constants \cite{jmgkou}. In particular, the measured $SU(3)$-splitting between $\pi$ and K decay constants,
\begin{equation}
 \frac{f_K}{f_\pi}\equiv1+\epsilon\label{decayratio}
\end{equation}
with $\epsilon=0.22\pm0.01$ of the order of $\left(m_K^2-m_\pi^2\right)/1\text{GeV}^2$, provides a rather interesting link between our present work on the $\eta-\eta'$ mixing and the so-called two-mixing-angle scheme high-lighted in ref.\cite{FEF}. Indeed, the equations
\begin{eqnarray}
 \theta_8 &=& \theta-\frac{2\sqrt2}{3}\epsilon\nonumber\\
 \theta_0 &=& \theta+\frac{2\sqrt2}{3}\epsilon
\end{eqnarray}
relate the universal mixing angle $\theta$ which diagonalizes the octet-singlet mass matrix (after renormalizing the meson fields) to the $\theta_{8,0}$ angles associated with the octet-singlet decay constants
\begin{eqnarray}
 f_8&=&\left(1+\frac{\epsilon}{3}\right)f_K\nonumber\\
 f_0&=&\left(1-\frac{\epsilon}{3}\right)f_K.
\end{eqnarray}
At $\mathscr{O}\left(p^2,0\right)$, $\epsilon=0$ and $\theta_8=\theta_0$ but $\theta$ cannot be determined. Yet, in this Letter, we have explicitly checked that the mixing angle 
\begin{equation}
 \theta_{th}\equiv-\frac{1}{2}\tan^{-1}\sqrt2\approx-27^\circ\label{angleConcl}
\end{equation}
which optimizes the $\eta-\eta'$ mass spectrum at lowest order is protected against quadratic one-loop divergences in the safe $m_\pi^2\rightarrow0$ limit. This result vindicates the approach based on Eq.\eqref{hierar} since $\theta_{th}$ is quite consistent with the physical mixing angle 
\begin{equation}
 \theta\approx-\left(22\pm1\right)^\circ
\end{equation}
directly extracted from the anomalous $J/\Psi\rightarrow\eta(\eta')\gamma$ decays \cite{jmgkou}. Indeed, higher order corrections are typically of the order of 20\%, as nicely illustrated in Eq.\eqref{decayratio}. In consequence, $\theta_8\approx-34^\circ$ and $\theta_0\approx-10^\circ$ within our specific momentum expansion supplemented by a large $N_c$ limit. However, any physical process only evaluated at the lowest order in the chiral expansion should rely on Eq.\eqref{angleConcl} if it involves on-shell or off-shell $\eta\left(\eta'\right)$, as it is the case in $\eta\left(\eta'\right)\rightarrow \gamma\gamma$ or in $K_L\rightarrow(\eta,\,\eta')\rightarrow\gamma\gamma$ decays, respectively.

\acknowledgments
One of us (C.D.) would like to thank Claude Duhr for adapting Feynrules to the QCD effective field theory. This work was supported by the Fonds National de la Recherche Scientifique and by the Belgian Federal Office for Scientific, Technical and Cultural Affairs through the Interuniversity Attraction Pole No. P6/11.

\end{document}